\documentclass[12pt]{article}%
\usepackage{graphicx}
\usepackage{amsmath}
\usepackage{amsfonts}
\usepackage{amssymb}%
\providecommand{\U}[1]{\protect\rule{.1in}{.1in}}
\newtheorem{theorem}{Theorem}

\begin{document}

\title{\textbf{Crystals in the Void \footnotetext{Part of this work has been carried
out in the framework of the Labex Archimede (ANR-11-LABX-0033) and of the
A*MIDEX project (ANR-11-IDEX-0001-02), funded by the \textquotedblleft
Investissements d'Avenir" French Government programme managed by the French
National Research Agency (ANR). Part of this work has been carried out at IITP
RAS. The support of Russian Foundation for Sciences (project No. 14-50-00150)
is gratefully acknowledged.}}}
\author{Senya Shlosman$^{\natural,\sharp,\flat}$\\$^{\natural}$Skolkovo Institute of Science and Technology, Moscow, Russia;\\$^{\sharp}$Aix Marseille University, University of Toulon, \\CNRS, CPT, Marseille, France;\\$^{\flat}$Inst. of the Information Transmission Problems,\\RAS, Moscow, Russia.\\Senya.Shlosman@cpt.univ-mrs.fr, shlos@iitp.ru}
\maketitle

\begin{abstract}
We study the problem of the crystal formation in vacuum.

\end{abstract}

\section{Introduction}

By now, the rigorous theory of the Wulff shapes is well developed in
statistical mechanics. It describes the asymptotic shape of a large (random)
droplet of one phase surrounded by the sea of coexisting different phase. Such
a large droplet arises if forced into a system at the phase coexistence by the
canonical constraint (see \cite{DKS, B, CP}), or else when it is created
spontaneously during the process of dynamical relaxation from the metastable
state to the stable one (\cite{SS}). All the papers cited are dealing with the
Ising model at zero magnetic field and the temperature below the critical one,
i.e. when there is the phase coexistence phenomenon.

In \cite{R} a different question is asked: suppose a certain given amount of
matter is put into void space (for example, this matter can be described by
the Ising model at a positive magnetic field, in which regime there is no
phase coexistence). Which shape this matter will assume? Alternatively,
consider a droplet of water, floating in oil, and then low the temperature so
that the water freezes. What will be the shape of the ice crystal one sees?

In the present note we provide rigorous answers to this question for some
models, which in some cases are different from the answers anticipated in
\cite{R}. We also formulate conjectures in situations where rigorous answers
are not available currently.

\section{Lattice case}

\subsection{Ising crystal in the void}

Consider the Ising $\pm1$ spins $\sigma$ in $d$ dimensions under magnetic
field $h>0$ at temperature $\beta^{-1},$ given by the Hamiltonian%
\[
H^{I}\left(  \sigma\right)  =-\sum_{x,y\ \mathrm{n.n.}}\sigma_{x}\sigma
_{y}-h\sum_{x}\sigma_{x},\ \ x,y\in\mathbb{Z}^{d}.
\]
The idea in \cite{R} is to consider the finite container $V$ as a third
parameter of the model. Namely, let $Z\left(  \beta,h,V\right)  $ denote the
partition function in $V$ with \textbf{free }boundary condition. Denote by
$\mathbb{T}_{N}^{d}$ the $d$-dimensional torus of volume $N^{d}.$ The
\textbf{void }crystal model is defined by the partition function%
\begin{equation}
\Xi\left(  \beta,h,N,\mathbf{f}\right)  =\sum_{\substack{V\subset
\mathbb{T}_{N}^{d}:\\\mathrm{vol}\left(  V\right)  =\left\lfloor
\mathbf{f}N^{d}\right\rfloor }}Z\left(  \beta,h,V\right)  . \label{10}%
\end{equation}
Here $\mathbf{f}>0$ is the number defining the proportion of the box
$\mathbb{T}_{N}^{d}$ filled by the Ising matter. We take $\mathbf{f}$ to be
small enough, so that the shape of the box $\mathbb{T}_{N}^{d}$ will have no
effect on the typical $V$-s.

The corresponding probability distribution $\mathbf{P}_{N}\left(  \ast\right)
$ on the boxes $V\subset\mathbb{T}_{N}^{d}$ with $\mathrm{vol}\left(
V\right)  =\left\lfloor \mathbf{f}N^{d}\right\rfloor $ is given by%
\begin{equation}
\mathbf{P}_{N}\left(  V\right)  =\frac{Z\left(  \beta,h,V\right)  }{\Xi\left(
\beta,h,N,\mathbf{f}\right)  }. \label{11}%
\end{equation}
This probability distribution is somewhat different from the one suggested in
\cite{R}: we do not restrict $V$ to be star-shaped, nor, in fact, that $V$ is
connected or simply-connected. This generality is physically more reasonable.

The question we want to study is the typical properties of the shapes $V$ as
the size $N$ of our system goes to $\infty.$ To this end we first reformulate
the question as a question about the behavior of a different model at the
phase coexistence.

The new model is again a lattice spin model $\eta$ on $\mathbb{Z}^{d},$ taking
now values $+1,0,-1,$ and defined by the Hamiltonian%
\begin{equation}
H\left(  \eta\right)  =-\sum_{x,y\ \mathrm{n.n.}}\eta_{x}\eta_{y}-h\sum
_{x}\eta_{x}-\left[  d+k\right]  \sum_{x}\left(  1-\eta_{x}^{2}\right)  .
\label{12}%
\end{equation}
In words, we put our $\pm1$ Ising model into the ideal gas of non-interacting
$0$ spins, which are subject to the magnetic field of strength $\left[
d+k\right]  .$ Note that the question of the behavior of the random boxes $V$
of size $\left\vert V\right\vert =\left\lfloor \mathbf{f}N^{d}\right\rfloor $
under the distribution $\left(  \ref{11}\right)  $ is equivalent to the
question of the behavior of the random boxes $W\left(  \eta\right)
\equiv\left\{  x\in\mathbb{T}_{N}^{d}:\eta_{x}\neq0\right\}  $ of the model
$\left(  \ref{12}\right)  ,$ considered

\begin{itemize}
\item at the temperature $\beta^{-1},$

\item in the same magnetic field $h>0,$

\item under `canonical' constraint $\sum_{x}\left(  1-\eta_{x}^{2}\right)
=\left(  1-\mathbf{f}\right)  N^{d},$

\item for any value of the magnetic field $\left[  d+k\right]  ,$ acting on
the $0$ spins.
\end{itemize}

Consider the translation-invariant ground state configurations of the
Hamiltonian $\left(  \ref{12}\right)  $. For $k<h$ it is the configuration
$\eta^{+}\equiv+1,$ for $k<h$ it is the configuration $\eta^{0}\equiv0,$ while
for $k=h$ the Hamiltonian $\left(  \ref{12}\right)  $ is degenerate and has
two ground state configurations: $\eta=+1\ $and $\eta=0.$ Moreover, it is easy
to check that at $k=h$ the \textit{Peierls} \textit{stability condition }holds:

Consider a finite box $V\subset\mathbb{Z}^{d},$ and define the configuration
$\eta^{V}$ by%
\[
\eta_{x}^{V}=\left\{
\begin{array}
[c]{cc}%
+1 & \text{if }x\in V,\\
0 & \text{otherwise.}%
\end{array}
\right.
\]
Then for some $\tau>0$ we have%
\[
H\left(  \eta^{V}\right)  -H\left(  \eta^{0}\right)  \geq\tau\left\vert
\partial V\right\vert ,
\]
where $\partial V=\left\{  \left\{  x,y\right\}  :x,y\ \mathrm{n.n.}%
\in\mathbb{Z}^{d},x\in V,y\notin V\right\}  .$ (In fact, $H\left(  \eta
^{V}\right)  -H\left(  \eta^{0}\right)  =\left\vert \partial V\right\vert ,$
so the Peierls constant $\tau$ can be taken to be $1.$)

Therefore the Pirogov-Sinai theory, see e.g. \cite{S}, applies to our model
$\left(  \ref{12}\right)  .$ It claims that for any $d\geq2$ there exists a
value $\beta_{d}<\infty$ such that for all $\beta>\beta_{d}$ there exist the
value $k\left(  \beta,h\right)  $ of the magnetic field $k$ at which there are
two translation-invariant Gibbs states corresponding to the Hamiltonian
$\left(  \ref{12}\right)  ;$ one is small perturbation of the configuration
$\eta^{+},\ $while the other -- of the configuration $\eta^{0}.$ Moreover,
$k\left(  \beta,h\right)  \rightarrow h,$ as $\beta\rightarrow\infty.$ (In
fact, $k\left(  \beta,h\right)  $ can be easily expressed via the free energy
$f\left(  \beta,h\right)  $ of the Ising model.)

Summarizing, we see that the study `in the void' -- of the behavior of the
random box $V$ under the distribution $\mathbf{P}_{N}\left(  V\right)  $
above, is equivalent to the study of the random box $W$ at the same
temperature $\beta$ and the field $k=k\left(  \beta,h\right)  ,$ i.e. at
coexistence (provided $\beta$ is large).

In particular, for $d=2$ all the machinery and all the results of \cite{DKS}
are valid in our situation, provided $\beta$ is large enough:

\begin{theorem}
\label{W} Let $h>0,$ $\beta>\beta_{d=2}$ and $0<\mathbf{f}<\frac{1}{10}.$
There exists the subset $\mathcal{V}_{N}$ of boxes, $\mathcal{V}_{N}=\left\{
V\subset\mathbb{T}_{N}^{2}:\mathrm{vol}\left(  V\right)  =\mathbf{f}%
N^{2}\right\}  ,$ such that $\mathbf{P}_{N}\left(  \mathcal{V}_{N}\right)
\rightarrow1$ as $N\rightarrow\infty,$ which has the following properties:

\textbf{1. }Among the connected components $\partial_{i}V$ of the boundary
$\partial V$ of a box $V\in\mathcal{V}_{N}$ there exists exactly one -- say,
$\partial_{1}V\equiv\Gamma\subset\partial V$ -- which is `big': $\mathrm{diam}%
\left(  \partial_{1}V\right)  \sim N;$ all other components are `small':
$\mathrm{diam}\left(  \partial_{i}V\right)  \sim\ln N,$ $i=2,3,...$

\textbf{2.} The contour $\Gamma$ has \textbf{asymptotic shape }$W,$ with
\begin{equation}
W=W\left(  \beta,h\right)  \subset\mathbb{R}^{2} \label{16}%
\end{equation}
being some smooth ($\mathcal{C}^{\infty}$) strictly convex centrally symmetric
closed curve. More precisely, for every $V\in\mathcal{V}_{N}$ and its big
boundary component $\Gamma=\partial_{1}V$ there exists a vector $x\left(
\Gamma\right)  $ such that the shifted contour, $\Gamma+x\left(
\Gamma\right)  ,$ satisfies
\begin{equation}
\mathrm{dist}_{\mathbf{H}}\left(  \Gamma+x\left(  \Gamma\right)  ,cNW\right)
\leq N^{3/4}. \label{15}%
\end{equation}
Here $\mathrm{dist}_{\mathbf{H}}$ is the Hausdorf distance, and the scaling
factor $c$ depends on $\beta,h$ and $\mathbf{f}$ only.

\textbf{3. }The curve $W$ is the Wulff shape, corresponding to the Hamiltonian
$\left(  \ref{12}\right)  $ at the temperature $\beta$ and magnetic fields
$h$, $k\left(  \beta,h\right)  .$ Its construction is explained below.
\end{theorem}

The proof of this theorem follows essentially the same lines as that in the
book \cite{DKS}. The difference is that \cite{DKS} treats the Ising model,
while here we have a different one. But the analysis of the proof given in
\cite{DKS} shows that its technique applies also to any 2D model within the
Pirogov-Sinai class. The differences are only notational and technical, though
they will result in the doubling of the length of the proof.

The situation in the 3D case is similar, though some details differ: instead
of the distance $\mathrm{dist}_{\mathbf{H}}$ one has to consider the $L^{1}%
$-distance, the exponent in $\left(  \ref{15}\right)  $ is not known (though
$\ln N$ is expected, compare with \cite{K}), the Wulff surface $W$ is not
strictly convex and is only $\mathcal{C}^{1},$ etc. For details, see \cite{B,
BIV, CP}.

\subsection{The Wulff shape}

In this section we will explain how to construct the curve $W$ entering our
theorem, see $\left(  \ref{16}\right)  .$ This curve is a solution to the
\textbf{Wulff variational problem}, stated below. The variational problem has
as its input a certain \textbf{surface tension function, }$\tau_{H,\beta
}\left(  \ast\right)  ,$ which is defined by the Hamiltonian $\left(
\ref{12}\right)  $ and the inverse temperature $\beta,$ and which will be
defined next.

\subsubsection{Wulff problem}

Wulff variational problem is formulated as follows. Let $\tau\left(
\mathbf{n}\right)  ,$ $\mathbf{n}\in S^{d-1}$ be some continuous function on
the unit sphere $S^{d-1}\subset\mathbb{R}^{d}$. We suppose that $\tau>0,$ and
that $\tau$ is even. For every closed compact (hyper)surface $M^{d-1}%
\subset\mathbb{R}^{d}\mathbb{\ }$we define its surface energy as%

\[
\mathcal{W}_{\tau}\left(  M\right)  =\int_{M}\tau\left(  \mathbf{n}%
_{s}\right)  ds,
\]
where $\mathbf{n}_{s}$ is the normal vector to $M$ at $s\in M.$ The functional
$\mathcal{W}_{\tau}\left(  M\right)  $ has the meaning of the surface energy
of the $M$-shaped droplet. It is called the \textit{Wulff functional. }Let
$W_{\tau}$ be the surface which minimizes $\mathcal{W}_{\tau}\left(
\cdot\right)  $ over all the surfaces enclosing the unit volume. Such a
minimizer does exist and is unique up to translation. It is called the
\textit{Wulff shape. }

The following is the geometric construction of $W_{\tau}.$ Consider the set%
\[
K_{\tau}=\left\{  \mathbf{x}\in\mathbb{R}^{d}:\forall\mathbf{n\in}%
S^{d-1}\mathbf{\;}\left(  \mathbf{x},\mathbf{n}\right)  \leq\,\tau\left(
\mathbf{n}\right)  \right\}  .
\]
If we define the half-spaces
\[
L_{\tau,\mathbf{n}}=\left\{  \mathbf{x}\in\mathbb{R}^{d}:\mathbf{\;}\left(
\mathbf{x},\mathbf{n}\right)  \leq\,\tau\left(  \mathbf{n}\right)  \right\}
,
\]
then
\begin{equation}
K_{\tau}=\cap_{\mathbf{n}}L_{\tau,\mathbf{n}}; \label{e1}%
\end{equation}
in particular, $K_{\tau}$ is convex. It turns out that%
\[
W_{\tau}=\lambda_{\tau}\partial\left(  K_{\tau}\right)  ,
\]
where the dilatation factor $\lambda_{\tau}$ is defined by the normalization:
$\mathrm{vol}\left(  \lambda_{\tau}K_{\tau}\right)  =1.$ The relation $\left(
\ref{e1}\right)  $ is called the \textit{Wulff construction}.

\subsubsection{Surface tension \label{st}}

In this subsection we specify the surface tension function $\tau=\tau
_{H,\beta}\left(  \ast\right)  ,$ which has to be used in the construction above.

Let $\theta\in\mathbb{S}^{1}$ be a unit vector in $\mathbb{R}^{2}.$ Let us
define the spin configuration $\eta^{\theta}$ on $\mathbb{Z}^{2}$ by%
\[
\eta_{x}^{\theta}=\left\{
\begin{array}
[c]{cc}%
+1 & \text{if }\left\langle x,\theta\right\rangle \geq0,\\
0 & \text{otherwise,}%
\end{array}
\right.
\]
and let us also put
\[
\eta_{x}^{+}\equiv+1,\ \ \eta_{x}^{0}\equiv0.
\]
Let $B_{n}\subset\mathbb{Z}^{2}$ be a square box centered at the origin, with
a side $2n.$ Consider the partition functions $Z\left(  \beta,h,k,B_{n}%
;\eta^{\theta}\right)  ,$ $Z\left(  \beta,h,k,B_{n};\eta^{+}\right)  ,$ and
$Z\left(  \beta,h,k,B_{n};\eta^{0}\right)  ,$ which are computed in the box
$B_{n}$ for the Hamiltonian $\left(  \ref{12}\right)  $ with boundary
conditions $\eta^{\theta},$ $\eta^{+}$ and $\eta^{0}.$ The surface tension
$\tau_{H,\beta}$ is defined as
\begin{equation}
\tau_{H,\beta}\left(  \theta\right)  =\lim_{n\rightarrow\infty}-\frac{1}{\beta
l\left(  \theta,n\right)  }\ln\frac{Z\left(  \beta,h,k,B_{n};\eta^{\theta
}\right)  }{\sqrt{Z\left(  \beta,h,k,B_{n};\eta^{+}\right)  Z\left(
\beta,h,k,B_{n};\eta^{0}\right)  }}, \label{17}%
\end{equation}
where $l\left(  \theta,n\right)  $ is the length of the segment, obtained by
intersecting the line $L\left(  \theta\right)  =\left\{  x\in\mathbb{R}%
^{2}:\left\langle x,\theta\right\rangle =0\right\}  $ and the box $B_{n}.$

\begin{theorem}
Let the field $k=k\left(  \beta,h\right)  .$ Then the limit $\left(
\ref{17}\right)  $ exists, is positive for $\beta$ large enough and is smooth
in $\theta.$ It also satisfies the `triangle inequality' (see relation (2.2.2)
in \cite{DKS}).
\end{theorem}

As a result, the properties of the curve $W,$ listed in the Theorem \ref{W},
follow, see again \cite{DKS}.

\section{Continuum case}

Here we consider the case of crystals in $\mathbb{R}^{d}.$ Much less is known
here rigorously.

We will treat point random fields, defined by the interaction $U\left(
x,y\right)  =U\left(  \left\vert x-y\right\vert \right)  ,$ which is supposed
to be superstable. For example, Lennard-Jones potential, or the potential%
\begin{equation}
U\left(  \left\vert x-y\right\vert \right)  =\left\{
\begin{array}
[c]{cc}%
+\infty & \text{if }\left\vert x-y\right\vert \leq1,\\
\left\vert x-y\right\vert -3 & \text{if }1<\left\vert x-y\right\vert <3,\\
0 & \text{if }\left\vert x-y\right\vert \geq3
\end{array}
\right.  \label{21}%
\end{equation}
of \cite{R} or just the hard-core interaction will go. The weight $w_{\beta
,z}\left(  \mathbf{x}\right)  $ of a configuration $\mathbf{x=}\left\{
x_{i}\in\mathbb{R}^{d},i=1,...,n\right\}  $ is given by%
\begin{equation}
w_{\beta,z}\left(  \mathbf{x}\right)  =z^{n}\exp\left\{  -\beta\sum
_{i<j}U\left(  x_{i},x_{j}\right)  \right\}  .\label{22}%
\end{equation}
The parameter $z>0$ is called \textit{activity.}

For a finite box $V\subset\mathbb{R}^{d}$ the partition function with free
boundary conditions is defined as%
\[
Z\left(  \beta,z,V\right)  \equiv Z\left(  \beta,z,V,\varnothing\right)
=\sum_{n=0}^{\infty}\int_{V^{n}}w_{\beta,z}\left(  \mathbf{x}\right)
\ d\mathbf{x.}%
\]

Since there is no natural measure on the space of all boxes $V\subset
\mathbb{T}_{N}^{d}$ (these notations refer now to the continuous case of
$\mathbb{R}^{d}$), we will proceed via some discretization procedure, the
effect of which vanishes in the thermodynamic limit. For every $N$ we consider
the partition of the torus $\mathbb{T}_{N}^{d}$ into cubes of size $\frac
{1}{N}$ -- i.e. into $N^{2d}$ cubes, and we call a box $V\subset\mathbb{T}%
_{N}^{d}$ an $N$-box iff $V$ is the union of these $\frac{1}{N}$ cubes. (An
$N$-box need not to be connected.)

The probability distribution $\mathbf{P}_{N}\left(  \ast\right)  $ on the
$N$-boxes $V$ of size $\left\lfloor \mathbf{f}N^{d}\right\rfloor $ that we
want to consider now is given by
\[
\mathbf{P}_{N}\left(  V\right)  =\frac{Z\left(  \beta,z,V\right)  }{\Xi\left(
\beta,z,N,\mathbf{f}\right)  },
\]
where the partition function%
\[
\Xi\left(  \beta,z,N,\mathbf{f}\right)  =\sum_{\substack{V\subset
\mathbb{T}_{N}^{d}:\\\mathrm{vol}\left(  V\right)  =\left\lfloor
\mathbf{f}N^{d}\right\rfloor }}Z\left(  \beta,z,V\right)
\]
is obtained by summing over all $N$-boxes. So we can proceed as in the
previous section, introducing the auxiliary non-interacting $\mathbf{y}$
particles, filling the complement $\mathbb{T}_{N}^{d}\setminus V$ and having
the activity $\zeta=\zeta\left(  z,\beta\right)  $ which brings them into
equilibrium with the $\mathbf{x}$-field, and try to apply the Wulff
construction in this situation.

\subsection{Surface tension: conjectures}

The first thing to be done is the definition of the surface tension. Contrary
to the Ising model case, which has one Gibbs state once $h>0,$ here the
situation is different, and it is reasonable to expect that when both $\beta$
and $z$ are large, our models have continuum of extremal Gibbs states. In the
3D case one expects the breaking of both the rotation and translation
symmetry, while in the 2D case the translation symmetry is not broken,
\cite{Ri}, and only rotation symmetry is expected to be broken. Therefore the
definition of the surface tension should include the choice of the pure phase.
For the hard core models defined above we take for the boundary condition
$\eta^{\uparrow}$ the centers of the densest lattice packing $\Pi^{d}$ of
balls of radius $\frac{\left(  1+\varepsilon\left(  \beta,z,d\right)  \right)
}{2}$ in $\mathbb{R}^{d}$, with one ball centered at the origin. The
orientation of the lattice is chosen in such a way that the intersection of
the packing $\Pi^{d}$ with the horizontal plane $\mathbb{R}^{d-1}%
\subset\mathbb{R}^{d}$ results in the packing $\Pi^{d-1}.$ The parameter
$\varepsilon\left(  \beta,z,d\right)  $ is chosen in such a way that the
density of points in the configuration $\eta^{\uparrow}$ coincides with the
density of particles in a Gibbs state defined by the weight $\left(
\ref{22}\right)  .$ In particular, $\varepsilon\left(  \beta,z,d\right)
\rightarrow0$ if $\beta\rightarrow\infty$ or if $z\rightarrow\infty.$
\footnote{In the initial version of the present paper the parameter
$\varepsilon\left(  \beta,z,d\right)  $ was absent. The idea to introduce it
is due to T. Richthammer.}

Similarly to the Section \ref{st}, for every $\theta\in\mathbb{S}^{d-1}$ we
introduce the point configuration $\eta^{\uparrow,\theta},$ which coincides
with $\eta^{\uparrow}$ in the half-space $\mathbb{R}_{\theta}=\left\{
x\in\mathbb{R}^{d}:\left\langle x,\theta\right\rangle \geq0\right\}  $ and
which is empty in the remaining half-space. Then we consider the partition
functions $Z\left(  \beta,z,\zeta\left(  z,\beta\right)  ,B_{n};\eta
^{\uparrow,\theta}\right)  ,$ $Z\left(  \beta,z,\zeta\left(  z,\beta\right)
,B_{n};\eta^{\uparrow}\right)  ,$ and $Z\left(  \beta,z,\zeta\left(
z,\beta\right)  ,B_{n};\varnothing\right)  $ in the cubic box $B_{n},$ and we
define%
\begin{equation}
\tau_{\beta,z}^{\uparrow}\left(  \theta\right)  =\lim_{n\rightarrow\infty
}-\frac{1}{\beta l\left(  \theta,n\right)  }\ln\frac{Z\left(  \beta
,z,\zeta\left(  z,\beta\right)  ,B_{n};\eta^{\uparrow,\theta}\right)  }%
{\sqrt{Z\left(  \beta,z,\zeta\left(  z,\beta\right)  ,B_{n};\eta^{\uparrow
}\right)  Z\left(  \beta,z,\zeta\left(  z,\beta\right)  ,B_{n};\varnothing
\right)  }},\label{20}%
\end{equation}
where, again, $l\left(  \theta,n\right)  $ is the measure of intersection of
the cube $B_{n}$ with the plane $\left\langle x,\theta\right\rangle =0.$ At
present, there is

\begin{enumerate}
\item no proof of existence of the limit function $\tau_{\beta,z}^{\uparrow
}\left(  \theta\right)  $,

\item no proof of positivity and non-trivial dependence of $\tau_{\beta
,z}^{\uparrow}\left(  \theta\right)  $ on $\theta$ for large $\beta$ and $z$.
\end{enumerate}

If we assume both, then it is safe to conjecture that the analog of the
theorem \ref{W} holds in the present situation, with the Wulff shape $W\left(
\beta,h\right)  $ replaced by $W^{\uparrow}\left(  \beta,z\right)  ,$ which is
the solution of the Wulff problem corresponding to the surface tension
$\tau_{\beta,z}^{\uparrow}.$ However, there is an important difference: in the
relation $\left(  \ref{15}\right)  ,$ instead of the shift $\Gamma+x\left(
\Gamma\right)  $ of the crystal $\Gamma$ one has to consider also the rotated
crystal, $\rho\left(  \Gamma\right)  \circ\left[  \Gamma+x\left(
\Gamma\right)  \right]  ,$ where $\rho\left(  \Gamma\right)  \in SO\left(
d\right)  .$ This extra rotation appears due to the choice of one of many
possible low temperature phases of our model, made in $\left(  \ref{20}%
\right)  $.

It seems that the simplest interaction for which the above conjectures
\textbf{1 }and \textbf{2 }can be proven in all dimensions $d\geq2$ is the one
given by $\left(  \ref{21}\right)  .$

\section{High temperature}

At high temperature (and low activity) we find ourselves in the uniqueness
regime, while the surface tension vanishes. As a result, no large crystal is
formed, i.e. all droplets are small. In the example of water bubble in oil it
means that water will be dispersed into infinitesimal droplets of no specific shape.

\textbf{Acknowledgement. }\textit{I thank the Department of Mathematics of the
University of Texas at Austin for its hospitality during my visit in May,
2017. I thank Professor Ch. Radin for the enlightening discussions of topics
treated in this paper.}

\end{document}